\documentclass[preprint,12pt]{elsarticle} 
\biboptions{numbers,sort&compress}
 
\usepackage[utf8]{inputenc} 
\usepackage[T1]{fontenc}    

\usepackage{amssymb}
\usepackage{lipsum}
\usepackage[nottoc]{tocbibind}

\usepackage{graphicx}
\usepackage{amssymb}   
\usepackage{amsfonts}
\usepackage{amsmath}
\usepackage{hyperref}
\usepackage{color}
\usepackage{overpic}
\usepackage{tikz}
\usepackage{braket}
\usepackage{rotating}
\usepackage{caption}
\usepackage{graphicx}
\usepackage{blindtext}
\usepackage{tcolorbox}
\usepackage{csquotes}
\usepackage{float}
\usepackage{setspace}
\usepackage{lineno}

\setcitestyle{square,comma}

\journal{Journal}

\begin{document}


\begin{frontmatter}



\title{Critical Iridium Demands arising from future Expansion of Proton Exchange Membrane Electrolysis}


\author[aff1,aff2]{B. Wortmann}
\author[aff1,aff2]{D. Stolten}
\author[aff1,aff3]{H. Heinrichs}

\address[aff1]{Forschungszentrum Jülich GmbH, Institute of Climate and Energy Systems, Jülich Systems Analysis (ICE-2), 52425 Jülich, Germany}
\address[aff2]{RWTH Aachen University, Chair for Fuel Cells, Faculty of Mechanical Engineering, 52062 Aachen, Germany}
\address[aff3]{University of Siegen, Professor for Energy Systems Analysis, Department of Mechanical Engineering, 57076 Siegen, Germany}

\begin{abstract}
Proton exchange membrane electrolysis (PEMEL) is a key technology for producing green hydrogen, but its scalability is limited by the use of scarce materials, particularly iridium. Iridium oxide, the preferred anode catalyst in PEMEL, offers exceptional stability but is produced only as a by-product of platinum mining, with annual output around 7.5 tons. This study estimates future iridium demand for PEMEL under various deployment scenarios and technological advances. Results show that meeting net zero targets will require both significant improvements in catalyst efficiency and access to roughly 30\% of global iridium production annually. Supply shortages could arise as early as 2030, earlier than previously anticipated. The analysis also reveals that long-term iridium needs beyond 2040 are significantly underestimated. These findings underscore the urgent need for innovation in material efficiency and recycling, and the importance of integrating resource constraints into energy policy and technology planning to ensure a sustainable hydrogen transition.
\end{abstract}




\end{frontmatter}



\section{Introduction}
\label{introduction}

Green hydrogen is essential for achieving a greenhouse gas neutral energy system. Its primary applications include fuel for heavy transport, chemical feedstock, energy storage, and combustion fuel in heavy industries \cite{HydrogenCouncil,APS,NZE,IRENA}. Hydrogen produced by water electrolysis using renewable electricity is particularly promising for meeting these diverse needs as it enables low carbon hydrogen production and supports energy system flexibility \cite{hydrorev}.

The commercially available water electrolysis technologies are Alkaline Electrolysis (AEL) and Proton Exchange Membrane Electrolysis (PEMEL) \cite{Rocha2024PEMlikeAWE}. While AEL offers modest cost advantages due to the absence of precious metal catalysts \cite{AEL}, PEMEL is expected to play a central role in future hydrogen production. Its higher power densities and superior load-following capabilities make it well-suited for integrating variable renewable energy sources \cite{Smolinka, Kiemel}. Moreover, PEMEL holds greater potential for innovation in catalyst efficiency, membrane durability, and scalability, unlike the more mature AEL technology \cite{Frauenhofer}.

Organizations such as the IEA, IRENA, and the Hydrogen Council have analyzed PEMEL scale-up, offering key projections for future hydrogen strategies. The IEA, for example, estimates a hydrogen demand of around 500 megatons by 2050 to meet net-zero targets \cite{IEAstudy}. Despite differences in scope, all studies project rapid electrolysis expansion, with PEMEL expected to account for about 40\% of capacity \cite{Smolinka}.

PEMEL technology depends on iridium as the anode catalyst, a critical and extremely scarce element with an annual global supply of around ~7.5 tons \cite{USGS}. No viable substitute currently exists, as iridium uniquely combines catalytic performance with corrosion resistance. This scarcity poses a potential bottleneck that could threaten decarbonization targets \cite{IRENAir,CHEREVKO2016170,BUTTLER20182440}. To mitigate this risk, recent research has focused on reducing iridium loadings and enhancing PEMEL performance \cite{Bernt_2018,Alia_2019,Möckl_2022,Bernt_2020}, though the extent to which iridium dependence can be lowered remains uncertain.

The role of iridium in PEMEL expansion has been explored through various projections of demand, supply constraints, and technological progress. One of the earliest studies, by Smolinka et al. \cite{Smolinka}, focused on Germany and concluded that, without major reductions in catalyst loading, national iridium demand could exceed global supply. However, the analysis was based only on optimistic assumptions about technological progress and recycling infrastructure. Kiemel et al. \cite{Kiemel} expanded on this by integrating criticality assessments with capacity projections and also warned of potential bottlenecks if German targets are overly ambitious. Both studies are geographically limited and do not account for global demand or competing sectors. Riedmeyer et al. \cite{Ir2} projected iridium needs under varying loadings and recycling rates but assumed an increasing iridium supply, which is unlikely given its status as a byproduct of platinum group metal (PGM) mining \cite{JMReport}. Minke et al. \cite{MINKE202123581} and Clapp et al. \cite{CLAPP2023114140} modeled scenarios with declining catalyst loadings. While Minke modeled a single optimistic trajectory, Clapp offered two scenarios and estimated that only 1.5 tons/year of iridium would be available for PEMEL, based on internal knowledge of the company Johnson Matthey. Despite their insights, none of these studies fully quantify how much of the iridium supply can be allocated to PEMEL given global market competition. Furthermore, the constructed models lack recursive calculation of iridium demands due to end-of-life replacements, leading to underestimations in demand projections. 

This study addresses key research gaps by developing a recursive model that estimates iridium demand from both initial PEMEL installations and end-of-life replacements, incorporating recycling effects. It also introduces refined supply scenarios by accounting for competing market demands and iridium price trends. Demand is assessed under two capacity expansion scenarios: one based on current real-world PEMEL projects \cite{IEAData}, and the other aligned with the IEA’s Net-Zero Emissions pathway \cite{NZE}.

\newpage

\section{Method}
\label{method}

This section outlines the method used to assess iridium demand, potential supply bottlenecks, and key sensitivities. It presents the core equations and model parameters varied across different scenarios.

\subsection{Modeling iridium demand}
\label{ss1}

The total primary iridium demand in year $i$ is defined as the sum of demand from capacity expansions $m_{cap}^i$ and end-of-life replacements ($m_{EOL}^i$), minus the recycled iridium available that year $m_{recycling}^i$ (equation (\ref{eq1})).

\begin{equation}
    m_{total}^i = m_{cap}^i + m_{EOL}^i - m_{recycling}^i
    \label{eq1}
\end{equation} 

The demand from PEMEL capacity expansion in year $i$, $m_{cap}^i$ is calculated as the product of the annual capacity additions $P_{el}^i$ and the iridium-specific power density $\omega^i$ $[kg \cdot GW^{-1}]$. The parameter $\omega^i$ reflects the iridium required to install one unit of capacity in year $i$.

\begin{equation}
    m_{cap}^i = P_{el}^i \cdot \omega^i
    \label{eq2}
\end{equation}

Based on this, the number of installed electrolyzers in year $i$ is calculated by dividing the annual capacity target by the average electrolyzer size of 1 MW \cite{REKSTEN202238106}. Each unit is assigned a lifetime drawn from normal distribution with mean $\tau$, a specified standard deviation $\sigma = 1/3 \tau$, and bounded lifetime limits $[1,2\cdot\tau]$. This reflects real life variability in operational conditions and manufacturing quality. Based on these lifetimes, end-of-life years are estimated to calculate both replacement demands $m_{EOL}^i$ and recycled iridium $m_{recycling}^i$. The required replacement capacity in year $i$ is derived from the capacity installed $\tau$ years earlier, denoted as $P^{i-\tau}_{el}$. The resulting replacements demand $m_{EOL}^i$ is given by equation \ref{eq3}. Newly installed units also receive sampled lifetimes, continuing the process recursively through the full analysis time horizon.

\begin{equation}
    m_{EOL}^i = P_{el}^{i-\tau} \cdot \omega^i
    \label{eq3}
\end{equation}

The recovered iridium quantity is calculated by multiplying the end-of-life scrap volume by a recycling efficiency factor $\gamma^i \in (0,1)$ which simultaneously reflects both the technical efficiency of the recycling process and the fraction of catalyst material that is successfully collected after use. This assumes negligible iridium loss during operational life. The resulting expression is given in equation (\ref{eq4}).

\begin{equation}
    m_{recycling}^i = (m_{cap}^{i-\tau} + m_{EOL}^{i-\tau}) \cdot \gamma
    \label{eq4}
\end{equation}

In summary, iridium demand is driven by four key parameters: iridium-specific power density $\omega^i$, electrolyzer lifetime $\tau$, capacity expansion $P^i_{el}$, and recycling rate $\gamma^i$. Their parametrization is detailed in the following subsections.

\subsection{Capacity expansion scenarios}
\label{ss2}

The first parameter for estimating iridium demand is the annual rate of PEMEL capacity installation. Two scenarios are used. The first is based on real and planned PEMEL projects compiled in the IEA database, which covers developments till 2030 \cite{IEAData} and is referred to as Business-As-Usual scenario (BAU). Capacity growth is extrapolated linearly to 2050, reaching 489 GW, nearly comparable to the IEA-APS scenario target of 580 GW, which reflects national climate pledges \cite{APS}. The second scenario is the IEA Net Zero Emissions (IEA-NZE) pathway \cite{NZE}, which targets net-zero emissions by 2050 and assumes a significantly larger role for PEMEL. Based on prior studies \cite{Smolinka}, PEMEL is projected to capture 40\% of the total electrolyzer market, resulting in 1468 GW by 2050 (see figure \ref{fig2}a).

The two scenarios differ substantially in their 2050 capacity targets. The BAU reflects a continuation of current trends, while the IEA-NZE scenario represents the effort required to meet net-zero goals. Comparing them reveals how primary and secondary iridium supply must scale to enable such ambitions.

\subsection{Iridium-specific power density and PEMEL lifetimes}
\label{ss3}

A key parameter for estimating iridium demand is the iridium-specific power density $\omega_i$ $[kg \cdot GW^{-1}]$, which reflects how efficiently iridium is utilized within a PEMEL stack. Literature values for $\omega^i$ vary widely from 0.34 to 2.0 $mgW^{-1}$ in 2024 and there is no consensus  on realistic benchmarks \cite{Smolinka,Babic_2017,CARMO20193450,Sparber, CLAPP2023114140}. Target values also differ, often lacking accompanying degradation data. As Clapp et al. note, lower iridium loadings tend to reduce stack lifetime, raising concerns about long-term viability \cite{CLAPP2023114140}. A single state of the art value of 0.65 $mgW^{-1}$ is currently considered the minimum to achieve a 10 year lifetime \cite{YU2020118194}. However, the relationship between the iridium-specific power density and lifetime remains uncertain due to ongoing technological developments and limited operational data. 
This study uses the approach of Clapp et al. \cite{CLAPP2023114140}, using exponentially decaying curves to model future iridium-specific power densities under two scenarios. The \textit{conservative} scenario assumes limited technological progress and slow reductions due to degradation constraints. The \textit{optimistic} scenario reflects accelerated improvements from advances in catalyst design (see \ref{fig2}b). These trajectories are used to estimate iridium demand under differing technological assumptions. Because lower iridium-specific power densities may affect system durability, model calculations also vary the average electrolyzer lifetime $\tau$ between 5 and 20 years to assess its impact on overall iridium demand.

\subsection{Recycling}
\label{ss4}

Reliable estimates of iridium recycling rates remain uncertain, as noted in prior studies \cite{MINKE202123581,CLAPP2023114140}. A useful benchmark has been drawn from platinum, a closely related PGM with similar recycling incentives. As highlighted by the Johnson Matthey PGM review \cite{JMReport}, platinum recycling rates in auto catalysts range from 50–70\%, despite occurring in open-loop systems, where materials are not returned to the original producer \cite{autocatalysts}. In contrast, iridium in PEMELs is recycled in closed-loop systems \cite{JMReport}, making a 70\% recycling rate a reasonable current estimate.

How iridium recycling could scale toward its technical maximum of 95–97\% \cite{Kiemel,CARMO20193450} remains uncertain. Given iridium’s high value and scarcity, manufacturers have strong incentives to improve recovery, particularly from end-of-life electrolyzers. Following Clapp et al. \cite{CLAPP2023114140}, this study assumes a linear increase in recycling efficiency from 70\% to 97\% by 2035 to reflect likely technological progress. Forecasting industrial developments of this scale involves significant uncertainties, particularly as companies often keep recycling technologies and capacities confidential leading to vast uncertainties in the estimation of accurate recycling rates. To account for these uncertainties, sensitivity calculations for different recycling rates are performed.

\subsection{Iridium supply estimates}
\label{ss5}

\begin{figure}[tbp]
    \includegraphics[width=10cm,height=16.0cm]{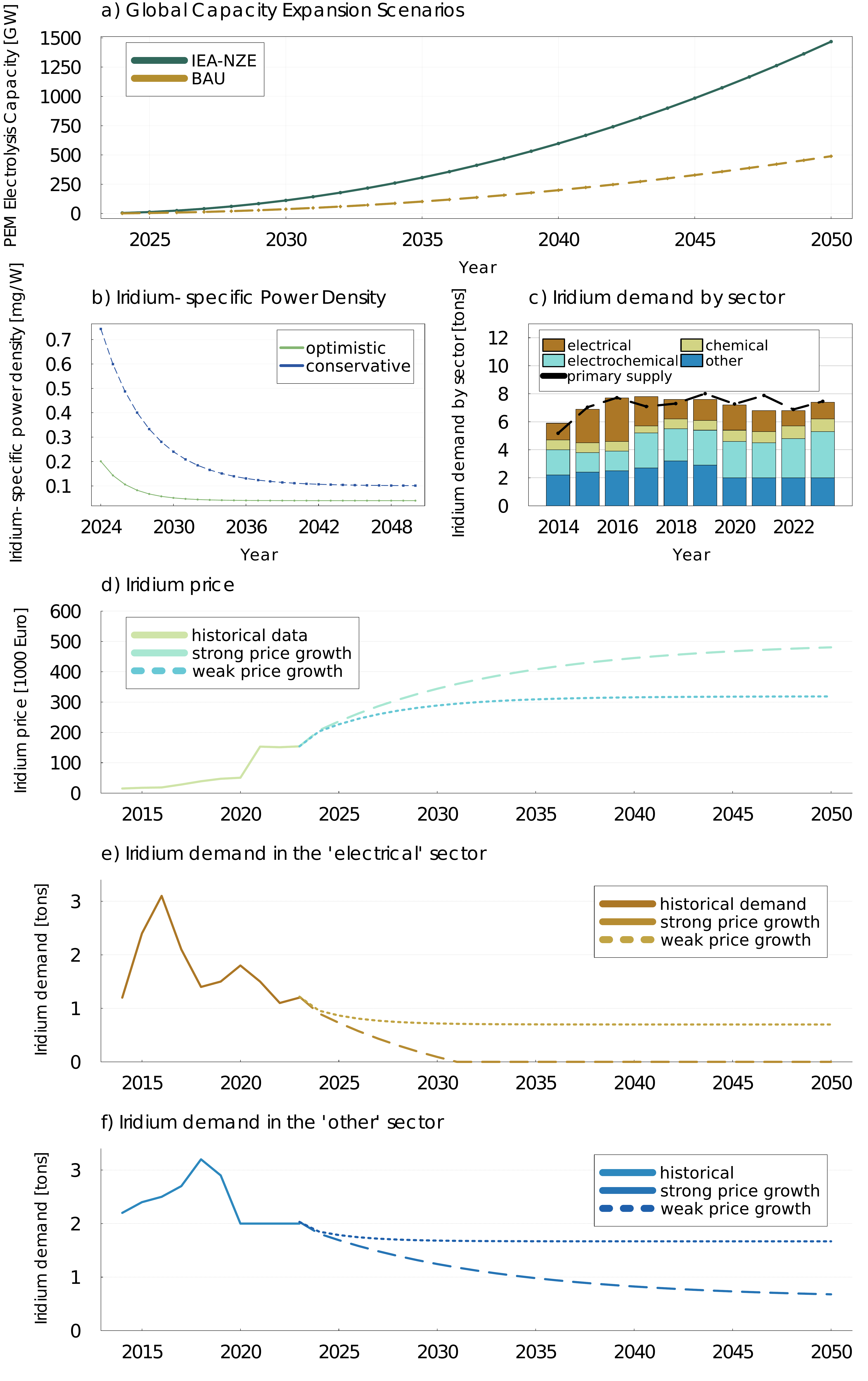}
\centering 
\caption{\textbf{a)} Capacity expansion for PEMEL up to 2050 \cite{IEAData,NZE}. \textbf{b)} Evolution of iridium-specific power density as in Clapp et al. \cite{CLAPP2023114140}. \textbf{c)} Supply and demand of iridium by sector from 2014 to 2023 \cite{JMReport} and supply \cite{USGS}. \textbf{d)} Iridium price forecast. \textbf{e)} Demand forecast "electrical" sector. \textbf{f)} Demand forecast "other" sector. Dampings are chosen such that a strong price growth ($\phi=0.9$) and a weak price growth ($\phi = 0.8$) is modeled.}
\label{fig2}
\end{figure}

Iridium is one of the rarest elements in Earth’s crust, with an annual primary production of about 7.5 tons \cite{USGS}. Its supply is highly volatile, influenced by factors such as mining strikes \cite{streiks}, and is dominated by South Africa, which accounts for 90–95\% of global output, followed by Russia, Canada, and Zimbabwe \cite{USGS}. As a byproduct of other PGM mining, its future production is difficult to project. Therefore, this study assumes a constant annual supply of 7.5 tons. Since the mid 2010s, production has mostly matched demand \cite{JMReport}, occasionally exceeding it and probably enabling limited stockpiling (see figure \ref{fig2}c).

Iridium demand spans four main sectors namely electrical, electrochemical, chemical, and others. Demand in the electrical sector is largely driven by applications like spark plugs and lithium crystal growth \cite{JMReport}. Spark plugs gained popularity in the early 2010s due to durability and low iridium prices, but recent price increases have led to growing substitution with platinum \cite{JMReport}.

Demand in the chemical sector has remained stable and is expected to persist \cite{JMReport}, as ruthenium–iridium catalysts play an irreplaceable role in key processes. Thus, this sector is not expected to free up iridium for PEMEL use. Iridium demand in the electrochemical sector has grown due to PEMEL deployment and copper foil production via electrolytic deposition. As copper foil is essential for lithium-ion batteries, demand for copper foil is expected to rise with battery expansion \cite{IEAbattery}. However, rising iridium prices may shift production toward rolling methods \cite{CopperFoil}, potentially easing iridium pressure and freeing supply for PEMEL.

The last sector of importance which is referred to as ”other” comprises a vast number of different kinds of products where among others, some examples are jewelry and applications in dental medicine \cite{JMReport}. 

To estimate iridium availability for PEMEL, we assess price-responsive demand in the “electrical” and “other” sectors, which historically show inverse trends with iridium prices \cite{Umicore}. These two sectors are considered because they are less dependent on iridium and offer substitution potential and show anti correlated behavior towards price movements, unlike the chemical and electrochemical sectors, where iridium use is often technologically essential. Price forecasts are generated using damped trend models that reduce the influence of older data over time, controlled by a damping factor $\phi$ \cite{Forecast} (see SI chapter 2). We model two cases: strong damping ($\phi=0.8$ and weak damping ($\phi=0.9$), corresponding to slower and faster price growth, respectively (see figure \ref{fig2}d-g). Values beyond $\phi=0.9$ are avoided, as they lead to extreme price spikes (e.g., €900,000/kg) and unrealistic demand collapse, particularly in the “other” sector. While demand declines are expected in the “electrical” sector due to available substitutes, the “other” sector, covering diverse, often price-insensitive products like jewelry is harder to predict. The chosen damping values ensure realistic outcomes across both sectors.

\newpage 

\section{Results}
\label{results}


\subsection{The conservative BAU Scenario}
\label{Projcon}

In the early years of capacity expansion, iridium demand rises sharply, peaking at approximately 2.1 tons in 2028. This is followed by a decline, reaching a local minimum of around 1.1 tons in 2037, driven by improvements in iridium-specific power density. Thereafter, demand steadily increases again, driven by accelerated capacity additions and growing replacement needs from end-of-life systems, eventually reaching a maximum of 3.1 tons.


\begin{figure*}[tbp]
    \includegraphics[width=\textwidth,height=14cm]{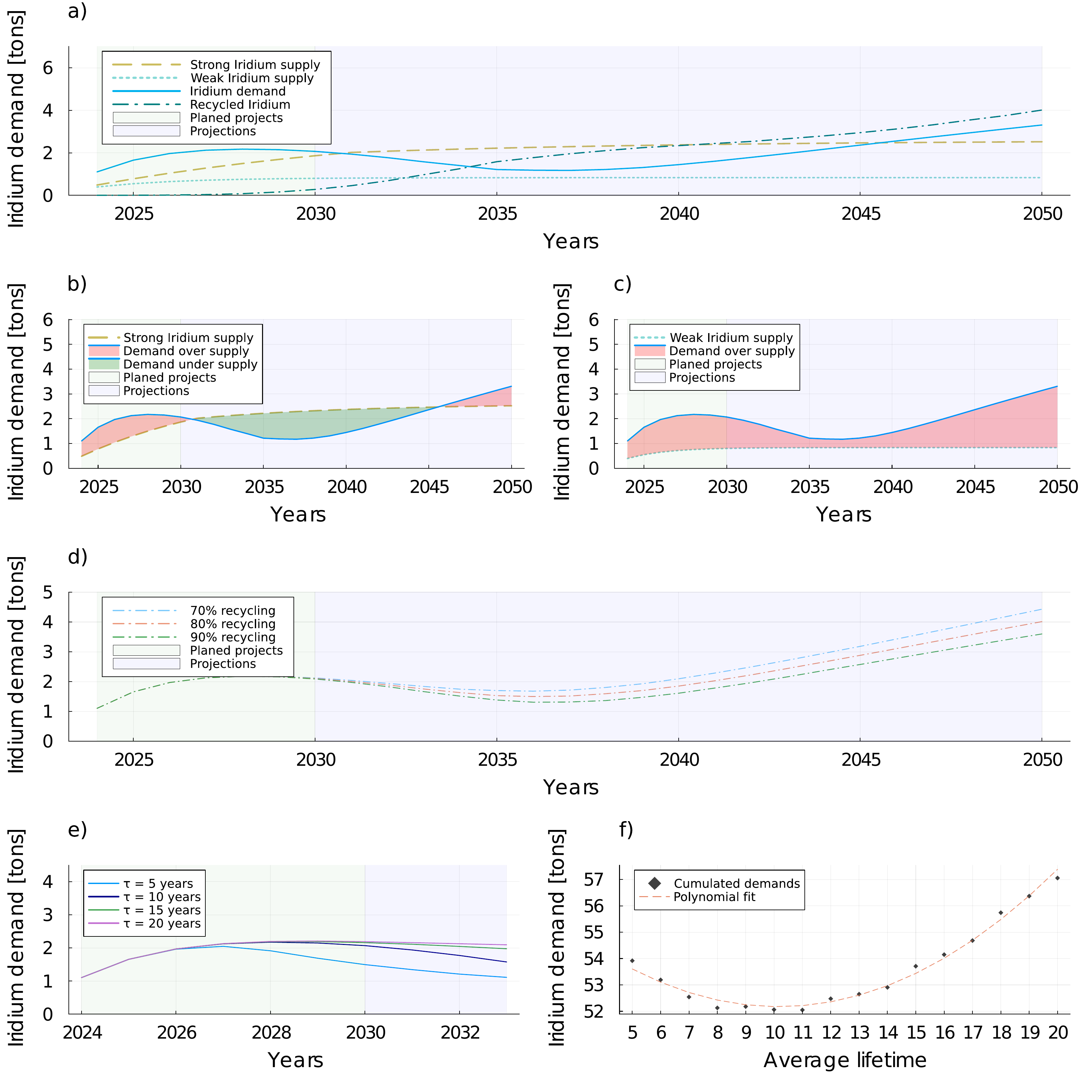}
\centering 
\caption{\textbf{The conservative BAU scenario:} Projected iridium demands with an average lifetime $\tau = 10$ years and recycling efficiency of 97\% by 2035 and corresponding supply projections. \textbf{a)} Demand projection, associated recycling curve and supply projections. \textbf{b)} Supply to demand gaps under strong supply. \textbf{c)} Supply to demand gaps for under weak supply. Red areas indicate time ranges in which demand is higher than supply whereas green areas indicate time ranges where supply is higher than demand. \textbf{d)} Iridium demands from variations of recycling rates $\gamma$ for 70\%, 80\% and 90\% reached by 2035. \textbf{e)} Iridium demands within the first 9 years from variations of the average lifetime of 5, 10, 15 and 20 years. \textbf{f)} Cumulative demands for the whole time range from lifetime variations ranging from 5 to 20 years.}
\label{Projection_con_all}
\end{figure*}

Figure \ref{Projection_con_all} b and c show the gaps between projected iridium demand and supply under the two supply scenarios. Red areas indicate periods where demand exceeds supply, while green areas represent potential stock accumulation when supply exceeds demand. These areas can be integrated to quantify additional iridium which is needed (red) or potentially storable (green).
Under the \textit{conservative} BAU scenario and the strong supply projection (Figure \ref{Projection_con_all}b), demand exceeds supply from the start of deployment until 2031, primarily driven by planned global PEMEL projects \cite{IEAData}. This suggests that some near-term projects may already face supply risks. The cumulative shortfall during this period amounts to roughly 4.5 tons, requiring an average 7.5\% increase in annual iridium supply to bridge the gap.
Between 2032 and 2046, however, supply exceeds demand, enabling the theoretical accumulation of approximately 10.2 tons of iridium in stock (without accounting for economic effects), sufficient to offset the 2.2-ton shortfall expected from 2047 onward. In contrast, under the weak supply projection (Figure \ref{Projection_con_all}f), demand exceeds supply throughout the entire time horizon. Bridging this gap would require an additional 30.2 tons of iridium, corresponding to an average annual supply increase of over 15\%.


Thus, under the weak supply projection (Figure \ref{Projection_con_all}c), increasing primary iridium production is the only viable way to realize the \textit{conservative} BAU scenario. However, this constraint does not apply under the strong supply projection (Figure \ref{Projection_con_all} b), where temporary supply shortages can be compensated through stock accumulation. 

Figure \ref{Projection_con_all}d-f illustrates the sensitivity of iridium demand to variations in average lifetime $\tau$ and recycling rate $\gamma$. As shown in Figure \ref{Projection_con_all}d, varying the ramp-up speed of recycling efficiency has little impact during the first six years of capacity expansion. However, differences grow over time, resulting in demand variations of up to 1 ton by 2050, which represent 13\% of today's primary iridium production. This underscores the long-term importance of achieving high recycling rates to meet capacity targets.

In contrast, changes in average lifetime $\tau$ have a more immediate impact. Figure \ref{Projection_con_all}e shows that shorter lifetimes lead to reduced primary iridium demand in the early years (2024–2033), as earlier end-of-life triggers earlier recycling. For instance, with $\tau=5$, the initial demand gap (Figure \ref{Projection_con_all}b) could be nearly halved to 2.5 tons, making it more feasible to bridge using existing stockpiles. 

However, this comes at a cost: shorter lifetimes increase the total number of replacements over time. Figure \ref{Projection_con_all}f shows cumulative demand rising with shorter $\tau$, with a minimum occurring around 10–11 years. Thus, while shorter lifetimes reduce near-term demand, they ultimately lead to higher cumulative material requirements over the whole time horizon. 

\subsection{The optimistic BAU Scenario}

\begin{figure*}[tbp]
    \includegraphics[width=\textwidth,height=14cm]{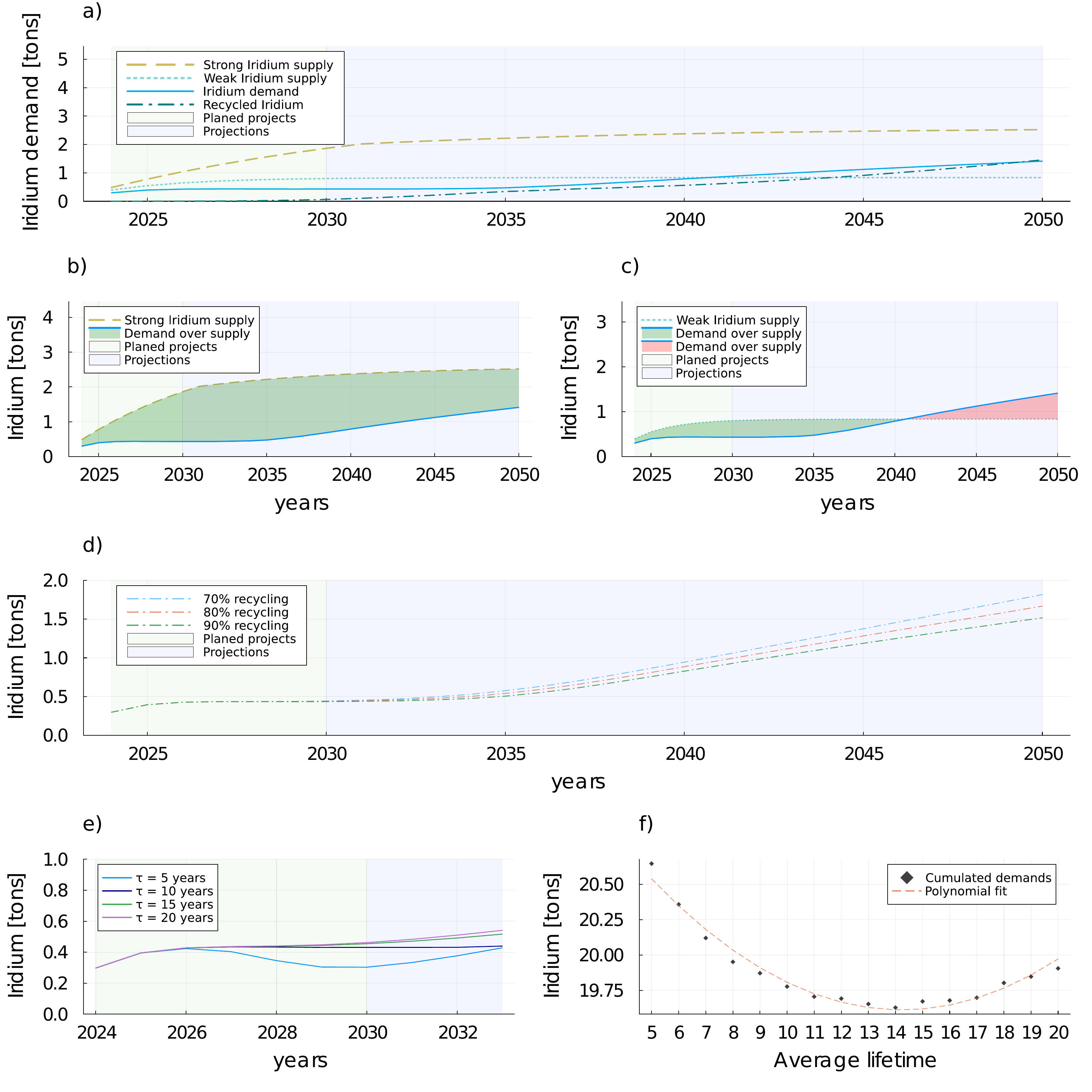}
\centering 
\caption{\textbf{The optimistic BAU scenario:} Projected iridium demands with an average lifetime $\tau = 10$ years and recycling efficiency of 97\% by 2035 and corresponding supply projections. \textbf{a)} Demand projection, associated recycling curve and supply projections. \textbf{b)} Supply to demand gaps under strong supply. \textbf{c)} Supply to demand gaps for under weak supply. Red areas indicate time ranges in which demand is higher than supply whereas green areas indicate time ranges where supply is higher than demand. \textbf{d)} Iridium demands from variations of recycling rates $\gamma$ for 70\%, 80\% and 90\% reached by 2035. \textbf{e)} Iridium demands within the first 9 years from variations of the average lifetime of 5, 10, 15 and 20 years. \textbf{f)} Cumulative demands for the whole time range from lifetime variations ranging from 5 to 20 years.}
\label{Projection_opt_all}
\end{figure*}

The BAU scenario under \textit{optimistic} iridium-specific power density assumptions shows significantly improved supply–demand dynamics. In contrast to the \textit{conservative} case, projected iridium demand only exceeds the weak supply projection after 2041 (Figure \ref{Projection_opt_all}c). The cumulative shortfall (red area) amounts to 3.1 tons, while the potential surplus (green area) reaches 4.6 tons, indicating that stockpiling during surplus years would be sufficient to meet capacity expansion targets.
Sensitivity analysis reveals that variations in the recycling rate (Figure \ref{Projection_opt_all}c) have negligible impact throughout the time horizon, suggesting that a 70\% recycling efficiency is sufficient, which is probably already achieved today \cite{CLAPP2023114140}. Additionally, lifetime variations indicate that an average lifetime of $\tau=14$ years minimizes cumulative demand (Figure \ref{Projection_opt_all}f), offering additional potential for material savings.
Overall, the \textit{optimistic} BAU scenario appears achievable under current recycling rates and a range of different electrolyzer lifetimes.

\subsection{The conservative IEA-NZE scenario}

\begin{figure*}[tbp]
    \includegraphics[width=\textwidth,height=14cm]{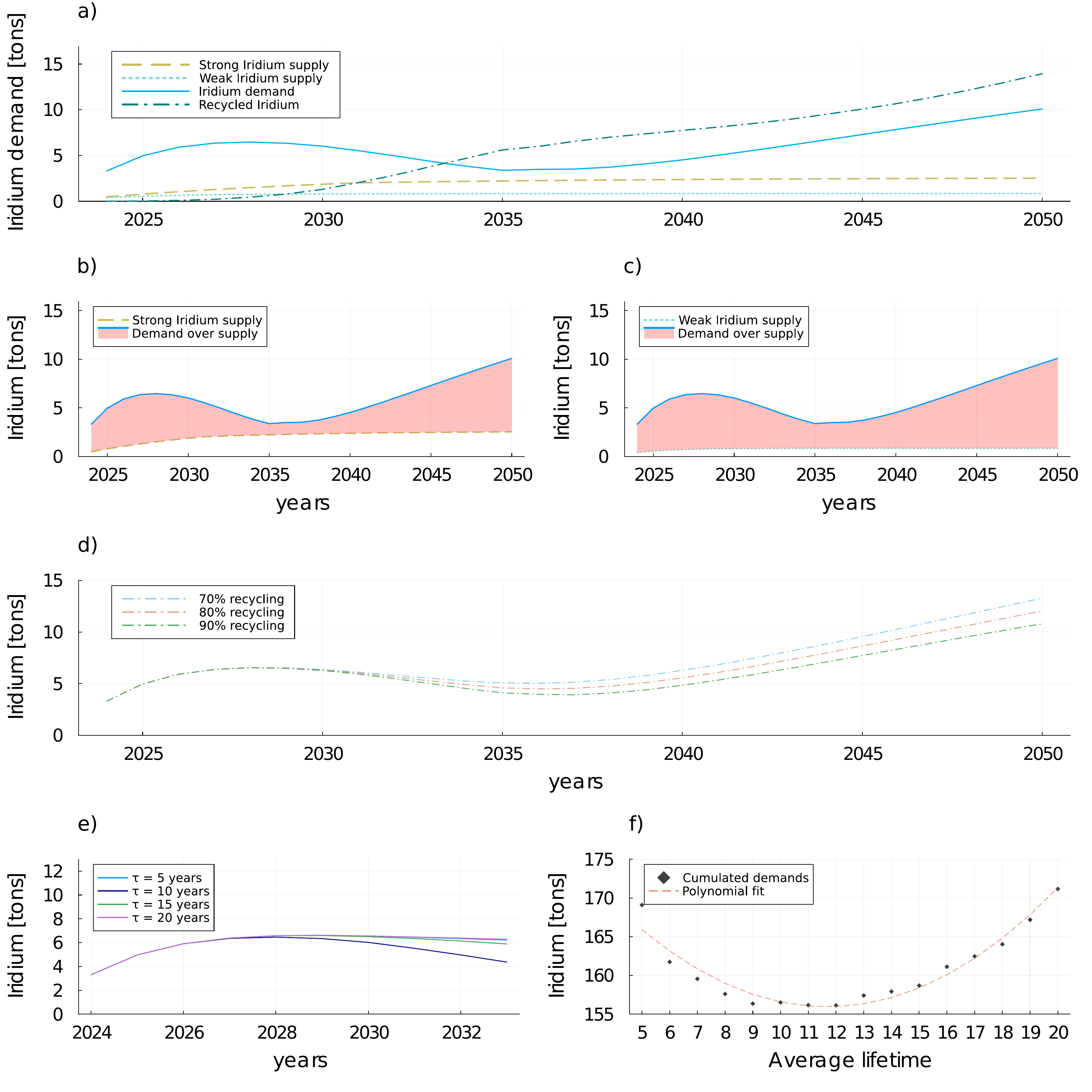}
\centering 
\caption{\textbf{The conservative IEA-NZE scenario:} Projected iridium demands with an average lifetime $\tau = 10$ years and recycling efficiency of 97\% by 2035 and corresponding supply projections. \textbf{a)} Demand projection, associated recycling curve and supply projections. \textbf{b)} Supply to demand gaps under strong supply. \textbf{c)} Supply to demand gaps for under weak supply. Red areas indicate time ranges in which demand is higher than supply whereas green areas indicate time ranges where supply is higher than demand. \textbf{d)} Iridium demands from variations of recycling rates $\gamma$ for 70\%, 80\% and 90\% reached by 2035. \textbf{e)} Iridium demands within the first 9 years from variations of the average lifetime of 5, 10, 15 and 20 years. \textbf{f)} Cumulative demands for the whole time range from lifetime variations ranging from 5 to 20 years.}
\label{NZE_con_all}
\end{figure*}

Compared to the BAU scenario, the IEA-NZE scenario features significantly more ambitious capacity expansion targets (Figure \ref{fig2}a), resulting in substantially higher projected iridium demand. As shown in Figure \ref{NZE_con_all}, demand exceeds both supply projections throughout the entire analysis period.

Bridging the supply gap in the \textit{conservative} IEA-NZE scenario would require approximately 101 tons of additional iridium under the strong supply projection, and around 135 tons under the weak supply projection. This corresponds to an average increase in primary iridium production of 49\% and 66\%, respectively. Sensitivity analyses of recycling rates and average lifetimes (Figure \ref{NZE_con_all}d-f) indicate that the magnitude of the shortfall remains largely unchanged. Even with a demand-minimizing lifetime of $\tau=11-12$. These results suggest that the \textit{conservative} IEA-NZE scenario is not feasible if annual primary iridium production remains fixed at approximately 7.5 tons.


\subsection{The optimistic IEA-NZE scenario}

\begin{figure*}[tbp]
    \includegraphics[width=\textwidth,height=14cm]{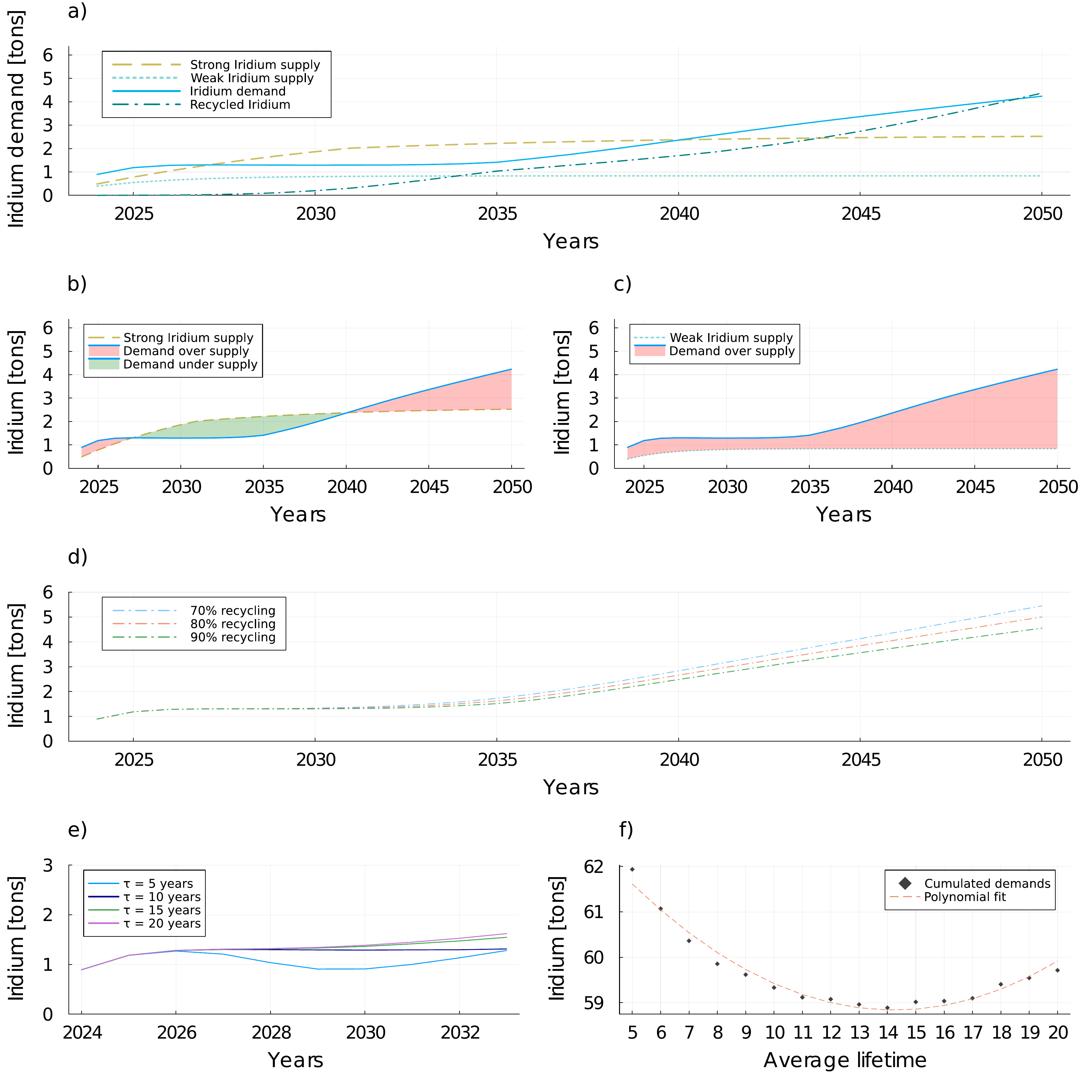}
\centering 
\caption{\textbf{The optimistic IEA-NZE scenario:} Projected iridium demands with an average lifetime $\tau = 10$ years and recycling efficiency of 97\% by 2035 and corresponding supply projections. \textbf{a)} Demand projection, associated recycling curve and supply projections. \textbf{b)} Supply to demand gaps under strong supply. \textbf{c)} Supply to demand gaps for under weak supply. Red areas indicate time ranges in which demand is higher than supply whereas green areas indicate time ranges where supply is higher than demand. \textbf{d)} Iridium demands from variations of recycling rates $\gamma$ for 70\%, 80\% and 90\% reached by 2035. \textbf{e)} Iridium demands within the first 9 years from variations of the average lifetime of 5, 10, 15 and 20 years. \textbf{f)} Cumulative demands for the whole time range from lifetime variations ranging from 5 to 20 years.}
\label{NZE_opt_all}
\end{figure*}

Projected iridium demand in the \textit{optimistic} IEA-NZE scenario is significantly lower than in the \textit{conservative} case (Figure \ref{NZE_opt_all}a). When compared to the strong supply projection, demand remains below supply from 2028 to 2040 (green area in Figure \ref{NZE_opt_all}b), with supply deficits occurring during 2024–2027 and again from 2040–2050 (red areas). The initial shortfall of 1.07 tons during 2024–2027 could likely be covered by existing iridium stocks, which can be estimated to be at least 1 ton based on Johnson Matthey data \cite{JMReport}. Between 2028 and 2040, a surplus of 6.9 tons could be accumulated, which would help offset the 9.7-ton deficit projected for 2040–2050, reducing the net gap to just 2.7 tons.
In contrast, under the weak supply projection (Figure \ref{NZE_opt_all}c), demand exceeds supply throughout the entire period, resulting in a cumulative shortfall of 38 tons.


In contrast to the \textit{optimistic} BAU scenario, appropriate speed in recycling ramping plays a significant role in the realization of the \textit{optimistic} IEA-NZE scenario (see figure \ref{NZE_opt_all}d-f). Comparing demand projections where the recycling rate stays consistently at 70\% (figure \ref{NZE_opt_all}d) to demand projections where it increases to 97\% (see figure \ref{NZE_opt_all}a)) a difference in demands of 9.2 tons arises by 2050. This emphasizes that high recycling efficiencies are crucial for the achievements of capacity expansion targets within the \textit{optimistic} IEA-NZE scenario.

Regarding the variations of lifetimes, similar to the scenarios before sensitivities yield that lower lifetimes $\tau$ lead to a decrease of iridium demands in the first years of capacity expansion (figure \ref{NZE_opt_all}e)). However, lower lifetimes again increase the cumulated iridium demands over the whole time range (figure \ref{NZE_opt_all}f)) as in any other scenario before. 

Since for the strong supply estimate, initial demands within the years 2024 to 2027 should not pose a problem due to the probable existence of sufficient iridium stocks (see chapter \ref{ss2}), exploitation of this effect is unnecessary. Calculations even suggest, that a lifetime of $\tau = 14$ would lead to a minimization of cumulated iridium demands. 

\newpage 

\newpage

\section{Discussion}
\label{SummaryDiscussion}

Under the BAU Scenario, capacity expansion targets are only at risk when both weak iridium supply projections and \textit{conservative} iridium-specific power density trajectories are assumed. In all other variations, the long-term targets appear achievable under average conditions. Notably, if iridium-specific power densities improve optimistically and strong supply projections hold, a recycling efficiency of just $\gamma = 0.7$ is sufficient to meet projected iridium demand. 
In contrast, realizing the IEA-NZE scenario requires a more favorable combination: \textit{optimistic} reductions in iridium-specific power density, strong iridium supply, and near-maximum recycling efficiency of 97\% by 2035. Even in the \textit{optimistic} case, a shortfall of 3.8 tons may arise, but this could potentially be covered by reallocating recycled iridium from other sectors or by utilizing potentially existing stockpiles.

\begin{figure}[tbp]
    \includegraphics[width=\textwidth]{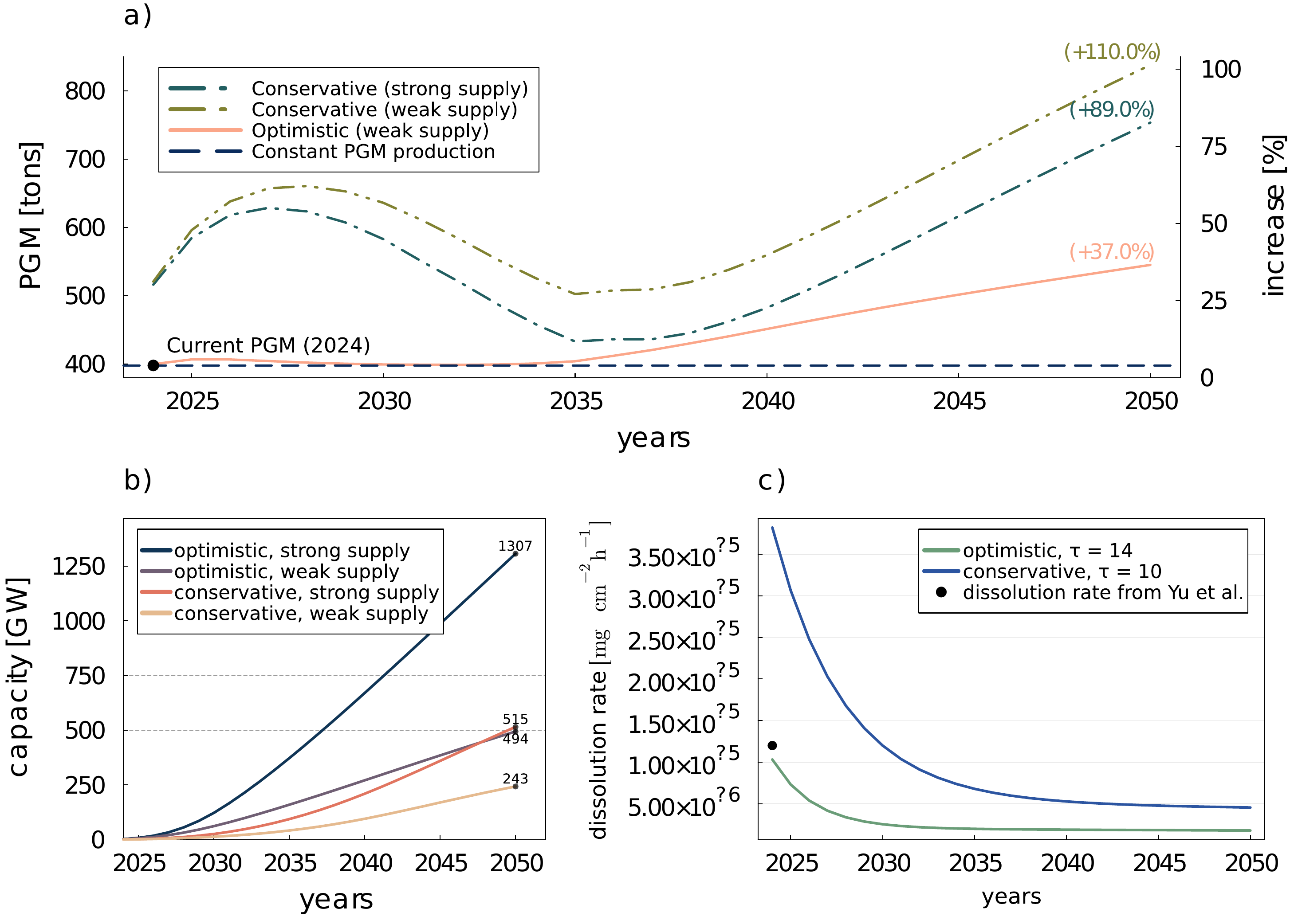}
\centering 
\caption{\textbf{a)} Required PGM production increase: Graph showing the required increase in PGM production in order to achieve sufficient iridium supply for the IEA-NZE scenario combinations listed in the legend. \textbf{b)} Maximum reachable capacities under constrained of iridium-specific power densities and supply projections. \textbf{c)} Required evolution of dissolution rates for $\tau = 10$ and $\tau = 14$.}
\label{PGMprodCapsDiss}
\end{figure}

All other scenario combinations would require an increase in primary iridium production. However, achieving this is far from straightforward. As previously noted, iridium is mined exclusively as a byproduct of other platinum group metals (PGMs) \cite{JMReport,CLAPP2023114140}, meaning its supply is governed by the demand for platinum and palladium, not iridium itself. To estimate the scale of required PGM production increases, it is assumed that iridium constitutes approximately 2\% of total PGM output \cite{USGS,Ir2}. Based on this ratio, the additional PGM extraction needed to meet iridium demand under the most challenging IEA-NZE scenarios is shown in Figure \ref{PGMprodCapsDiss} a.


Estimates indicate that in both \textit{conservative} scenarios, PGM production would need to increase by at least 25\% as early as 2024 to generate sufficient iridium to meet projected demand. In the \textit{optimistic} scenario, a more gradual increase starting around 2035 would be required. However, it remains uncertain whether primary PGM production will grow at all. For example, approximately 40\% of mined platinum is currently used in automotive catalytic converters \cite{autocatalysts}, a demand segment expected to decline with the ongoing electrification of vehicle fleets.

Conversely, several studies suggest that overall platinum demand may continue to rise due to its broad range of applications, including jewelry, glass manufacturing, electrolysis, and fuel cells \cite{JMReport,Pt}. In particular, the latter is expected to play a critical role in sustaining PGM production, highlighting the interdependence of different platinum group metals. Reverdiau et al. \cite{REVERDIAU202139195} estimate that platinum demand from PEMEL fuel cells alone could reach 1,000 tons annually by 2050, which would correspond to more than a five-fold increase compared to 2024 \cite{WPIC2025Q4}. Under such conditions, the IEA-NZE targets could remain achievable, provided iridium-specific power densities improve optimistically. However, the \textit{conservative} IEA-NZE scenarios would still be unattainable, as their initial material demands exceed feasible supply under any circumstance. This is further illustrated in Figure \ref{PGMprodCapsDiss}b, which shows maximum achievable capacity expansion under both \textit{conservative} and \textit{optimistic} assumptions for iridium-specific power density and supply, based on Equation (\ref{caps}).

\begin{equation}
    P_{el}^{i} = \frac{\omega^i}{m_{total}^i}
    \label{caps}
\end{equation}

Thus, iridium-specific power density $\omega^i$ emerges as the most critical factor influencing iridium demand from PEMEL electrolysis. As previously discussed, literature values for current $\omega^i$ vary significantly \cite{CLAPP2023114140}. Eikeng et al. report a state-of-the-art value of 750 $[kg \cdot GW^{-1}]$ \cite{EIKENG2024433}, aligning with the starting point of the \textit{conservative} scenario. If this value reflects current practice, achieving the IEA-NZE targets would require a rapid and substantial reduction in $\omega^i$.
While a 90\% reduction is considered technically feasible, comparable to past platinum reductions in PEMEL fuel cells \cite{Kiemel,EIKENG2024433,CLAPP2023114140,doi:10.1021/acs.jpclett.6b00216}, the decisive factor is the speed at which such improvements can be realized. Reducing $\omega^i$ early is essential not only to meet short-term capacity targets but also to ensure that iridium enters the recycling loop soon enough to reduce reliance on primary supply in later years. To support the IEA-NZE scenario, at least a threefold decrease, from 750 to 250 $[kg \cdot GW^{-1}]$ would be needed well before 2030. However, given current uncertainties in performance data, the likelihood of achieving such reductions remains difficult to assess. 


As highlighted by Clapp et al. \cite{CLAPP2023114140}, reducing dissolution rates which quantifies the degradation of the iridium at the anode (measured in $mg \cdot cm^2 \cdot h^{-1}$) is essential for lowering iridium-specific power densities while maintaining sufficient lifetimes. Figure \ref{PGMprodCapsDiss}c presents the required dissolution rates to sustain 10- and 14-year electrolyzer lifetimes under both \textit{conservative} and \textit{optimistic} $\omega^i$ trajectories. It also includes the current state-of-the-art dissolution rate for a 10-year lifetime, as reported by Yu et al. \cite{YU2020118194}.
The comparison shows that the dissolution rates required under \textit{conservative} assumptions are substantially higher than those already demonstrated by Yu et al., suggesting that a significant reduction in $\omega^i$ while maintaining a 10-year lifetime is likely achievable in the near term. However, achieving 14-year lifetimes at similarly low iridium loadings remains uncertain and would require further technological advances.

If iridium-specific power densities do not decline rapidly enough, one remaining, but questionable, strategy to reduce initial iridium demand would be to artificially shorten PEMEL lifetimes. As shown in the sensitivity analysis in Section \ref{results}, setting the average lifetime to $\tau=5$ years can significantly lower primary iridium demand in the early years by accelerating entry into the recycling loop. For example, if iridium requirements per unit of capacity were reduced from 750 to 325 $[kg \cdot GW^{-1}]$ , the recycled iridium from older electrolyzers could nearly support the installation of twice as many new units. However, it remains uncertain whether such an approach would be economically viable without external support, such as targeted subsidies.

\newpage

\section{Conclusion}
\label{Conclusion}

This study aimed to quantify future iridium demand from PEM electrolysis capacity expansions using a mathematical model that integrates deployment targets, iridium-specific power densities, electrolyzer lifetimes, recycling rates, and supply projections. Two scenarios were formulated for capacity expansions, iridium-specific power densities and iridium supplies to capture a range of possible futures, and  sensitivities based on lifetime and recycling variations were evaluated accordingly.
While actual demand trajectories will likely fall between the bounds of the different scenarios, the analysis clearly shows that meeting ambitious capacity targets, especially those aligned with the IEA-NZE scenario, requires a substantial and extremely fast reduction in iridium-specific power densities, along with either increased primary supply or high-efficiency recycling. Even so, achieving 40\% PEMEL market share appears infeasible under \textit{conservative} assumptions but remains within reach under more \textit{optimistic} conditions (see Figure \ref{PGMprodCapsDiss}b).
In cases where iridium bottlenecks limit PEM deployment, alternative electrolysis technologies offer viable pathways. Anion Exchange Membrane (AEM) electrolysis and Alkaline Electrolysis (AEL) could be used in applications where high dynamic response is not essential, allowing for strategic technology diversification. Careful techno-economic matching between electrolysis type and operational context could thus reduce reliance on PEMEL and alleviate critical material constraints.
Additional mitigation strategies include the development of novel catalysts, such as iridium–ruthenium alloys \cite{EIKENG2024433}, and the recovery of iridium from PGM tailings \cite{GIBSON2023108216}. Another avenue, not yet explored in the literature, involves the deliberate reduction of PEMEL lifetimes by lowering catalyst loading. While this would increase replacement frequency, it could reduce initial iridium demand and accelerate entry into the recycling loop, thereby supporting faster capacity deployment under supply constraints. The feasibility of such a strategy, however, warrants further economic and technical investigation. Beyond that, substitution of iridium with alternative materials in other industries represents an additional lever to alleviate criticality pressures. Continued innovation in catalyst and material science could reduce iridium dependence in non-hydrogen applications, indirectly improving the resource outlook for PEM electrolysis.

Overall, this study underscores the urgency of integrating material availability into electrolysis deployment strategies and highlights the importance of continued innovation in catalyst design, recycling systems, and technology choice to support sustainable hydrogen production.







\section*{Declaration of Competing Interest}
The authors declare that they have no known competing financial interests or personal relationships that could have influenced the work reported in this paper.

\section*{Declaration of generative AI}
During the preparation of this work the author used ChatGPT and DeepL Write in order to improve language and readability. After using this tool, the author reviewed and edited the content as needed and takes full responsibility for the content of the publication

\section*{Data Availability Statement}
The data that supports the findings of this study are available in the supplementary material of this article.

\section*{Acknowledgments}
This work was funded by the European Union (ERC, MATERIALIZE, 101076649). Views and opinions expressed are however those of the authors only and do not necessarily reflect those of European Union or the European Research Council Executive Agency. Neither the European Union nor the granting authority can be held responsible for them. Furthermore, we would like to thank Mark Clapp and Margery Ryan from Johnson Matthey PGM Market Research for fruitful discussion. 


\bibliographystyle{elsarticle-num} 
\bibliography{references}

\begin{thebibliography}{10}
\expandafter\ifx\csname url\endcsname\relax
  \def\url#1{\texttt{#1}}\fi
\expandafter\ifx\csname urlprefix\endcsname\relax\def\urlprefix{URL }\fi
\expandafter\ifx\csname href\endcsname\relax
  \def\href#1#2{#2} \def\path#1{#1}\fi

\bibitem{HydrogenCouncil}
H.~Council, M.~Company, Hydrogen for net-zero: A critical costcompetitive energy vector, \url{https://hydrogencouncil.com/wpcontent/uploads/2021/11/Hydrogen-for-Net-Zero.pdf}, accessed: 2024-01-16 (2021).

\bibitem{APS}
P.~IEA, Announced pledges scenario, \url{https://www.iea.org/reports/global-energy-and-climate-model}, accessed: 2024-04-16 (2023).

\bibitem{NZE}
P.~IEA, Net zero emissions scenario, \url{https://www.iea.org/reports/global-energy-and-climate-model}, accessed: 2024-04-16 (2023).

\bibitem{IRENA}
E.~Taibi, H.~Blanco, R.~Miranda, M.~Carmo, Irena: Green hydrogen cost reduction: Scaling up electrolysers to meet the 1.50c climate goal, \url{https://irena.org/-/media/Files/IRENA/Agency/Publication/2020/Dec/IRENA_Green_hydrogen_cost_2020.pdf.} (2020).

\bibitem{hydrorev}
M.~Nasser, T.~Megahed, S.~Ookawara, H.~Hassan, A review of water electrolysis–based systems for hydrogen production using hybrid/solar/wind energy systems, Environmental Science and Pollution Research 29 (10 2022).
\newblock \href {https://doi.org/10.1007/s11356-022-23323-y} {\path{doi:10.1007/s11356-022-23323-y}}.

\bibitem{Rocha2024PEMlikeAWE}
F.~Rocha, C.~Georgiadis, K.~V. Droogenbroek, R.~Delmelle, X.~Pinon, G.~Pyka, G.~Kerckhofs, F.~Egert, F.~Razmjooei, S.-A. Ansar, S.~Mitsushima, J.~Proost, \href{https://doi.org/10.1038/s41467-024-51704-z}{Proton exchange membrane-like alkaline water electrolysis using flow-engineered three-dimensional electrodes}, Nature Communications 15 (2024) 7444.
\newblock \href {https://doi.org/10.1038/s41467-024-51704-z} {\path{doi:10.1038/s41467-024-51704-z}}.
\newline\urlprefix\url{https://doi.org/10.1038/s41467-024-51704-z}

\bibitem{AEL}
F.~Gambou, D.~Guilbert, M.~Zasadzinski, H.~Rafaralahy, A comprehensive survey of alkaline electrolyzer modeling: electrical domain and specific electrolyte conductivity, \url{https://hal.science/hal-03663132/document}, [Online; accessed 28.03.2024] (2024).

\bibitem{Smolinka}
T.~Smolinka, N.~Wiebe, P.~Sterchele, A.~Palzer, F.~Lehner, M.~Jansen, K.~Steffen, R.~Miehe, S.~Wahren, F.~Zimmermann, Industrialisierung der wasser elektrolyse in deutschland: chancen und herausforderungen für nachhaltigen wasserstoff für verkehr, strom und waerme, Frauenhofer (2018).

\bibitem{Kiemel}
K.~Steffen, S.~Tom, L.~Franz, F.~Johannes, S.~Alexander, M.~Robert, \href{https://onlinelibrary.wiley.com/doi/abs/10.1002/er.6487}{Critical materials for water electrolysers at the example of the energy transition in germany}, International Journal of Energy Research 45~(7) (2021) 9914--9935.
\newblock \href {http://arxiv.org/abs/https://onlinelibrary.wiley.com/doi/pdf/10.1002/er.6487} {\path{arXiv:https://onlinelibrary.wiley.com/doi/pdf/10.1002/er.6487}}, \href {https://doi.org/https://doi.org/10.1002/er.6487} {\path{doi:https://doi.org/10.1002/er.6487}}.
\newline\urlprefix\url{https://onlinelibrary.wiley.com/doi/abs/10.1002/er.6487}

\bibitem{Frauenhofer}
M.~Holst, S.~Aschbrenner, T.~Smolinka, C.~Voglstätter, G.~Grimm, Cost forecast for low-temperature electrolysis – technology driven bottom-up prognosis for pem and alkaline water electrolysis systems, \url{https://www.ise.fraunhofer.de/content/dam/ise/de/documents/presseinformationen/2022/2021-11-17_CATF_Report_Electrolysis_final.pdf?utm_source=chatgpt.com}, [Online; accessed 16.12.2024] (2021).

\bibitem{IEAstudy}
I.~E. Agency, Global hydrogen review, \url{https://www.iea.org/reports/global-hydrogen-review-2021}, [Online; accessed 28.03.2024] (2022).

\bibitem{USGS}
U.~S.~G. Survey, Platinum-group-metals survey, \url{https://www.usgs.gov/centers/national-minerals-information-center/platinum-group-metals-statistics-and-information}, accessed: 2024-01-16.

\bibitem{IRENAir}
IRENA, Global renewables outlook: Energy transformation 2050, \url{https://www.irena.org/publications/2020/Apr/Global-Renewables-Outlook-2020}, [Online; accessed 28.03.2024] (2020).

\bibitem{CHEREVKO2016170}
S.~Cherevko, S.~Geiger, O.~Kasian, N.~Kulyk, J.-P. Grote, A.~Savan, B.~R. Shrestha, S.~Merzlikin, B.~Breitbach, A.~Ludwig, K.~J. Mayrhofer, \href{https://www.sciencedirect.com/science/article/pii/S0920586115004940}{Oxygen and hydrogen evolution reactions on ru, ruo2, ir, and iro2 thin film electrodes in acidic and alkaline electrolytes: A comparative study on activity and stability}, Catalysis Today 262 (2016) 170--180, electrocatalysis.
\newblock \href {https://doi.org/https://doi.org/10.1016/j.cattod.2015.08.014} {\path{doi:https://doi.org/10.1016/j.cattod.2015.08.014}}.
\newline\urlprefix\url{https://www.sciencedirect.com/science/article/pii/S0920586115004940}

\bibitem{BUTTLER20182440}
A.~Buttler, H.~Spliethoff, \href{https://www.sciencedirect.com/science/article/pii/S136403211731242X}{Current status of water electrolysis for energy storage, grid balancing and sector coupling via power-to-gas and power-to-liquids: A review}, Renewable and Sustainable Energy Reviews 82 (2018) 2440--2454.
\newblock \href {https://doi.org/https://doi.org/10.1016/j.rser.2017.09.003} {\path{doi:https://doi.org/10.1016/j.rser.2017.09.003}}.
\newline\urlprefix\url{https://www.sciencedirect.com/science/article/pii/S136403211731242X}

\bibitem{Bernt_2018}
M.~Bernt, A.~Siebel, H.~A. Gasteiger, \href{https://dx.doi.org/10.1149/2.0641805jes}{Analysis of voltage losses in pem water electrolyzers with low platinum group metal loadings}, Journal of The Electrochemical Society 165~(5) (2018) F305.
\newblock \href {https://doi.org/10.1149/2.0641805jes} {\path{doi:10.1149/2.0641805jes}}.
\newline\urlprefix\url{https://dx.doi.org/10.1149/2.0641805jes}

\bibitem{Alia_2019}
S.~M. Alia, S.~Stariha, R.~L. Borup, \href{https://dx.doi.org/10.1149/2.0231915jes}{Electrolyzer durability at low catalyst loading and with dynamic operation}, Journal of The Electrochemical Society 166~(15) (2019) F1164.
\newblock \href {https://doi.org/10.1149/2.0231915jes} {\path{doi:10.1149/2.0231915jes}}.
\newline\urlprefix\url{https://dx.doi.org/10.1149/2.0231915jes}

\bibitem{Möckl_2022}
M.~Möckl, M.~F. Ernst, M.~Kornherr, F.~Allebrod, M.~Bernt, J.~Byrknes, C.~Eickes, C.~Gebauer, A.~Moskovtseva, H.~A. Gasteiger, \href{https://dx.doi.org/10.1149/1945-7111/ac6d14}{Durability testing of low-iridium pem water electrolysis membrane electrode assemblies}, Journal of The Electrochemical Society 169~(6) (2022) 064505.
\newblock \href {https://doi.org/10.1149/1945-7111/ac6d14} {\path{doi:10.1149/1945-7111/ac6d14}}.
\newline\urlprefix\url{https://dx.doi.org/10.1149/1945-7111/ac6d14}

\bibitem{Bernt_2020}
M.~Bernt, J.~Schröter, M.~Möckl, H.~A. Gasteiger, \href{https://dx.doi.org/10.1149/1945-7111/abaa68}{Analysis of gas permeation phenomena in a pem water electrolyzer operated at high pressure and high current density}, Journal of The Electrochemical Society 167~(12) (2020) 124502.
\newblock \href {https://doi.org/10.1149/1945-7111/abaa68} {\path{doi:10.1149/1945-7111/abaa68}}.
\newline\urlprefix\url{https://dx.doi.org/10.1149/1945-7111/abaa68}

\bibitem{Ir2}
R.~Riedmayer, B.~Paren, L.~Schofield, Y.~Shao-Horn, D.~Mallapragada, Proton exchange membrane electrolysis performance targets for achieving 2050 expansion goals constrained by iridium supply, Energy and Fuels 37~(12) (2023) 8614--8623, publisher Copyright: {\textcopyright} 2023 American Chemical Society.
\newblock \href {https://doi.org/10.1021/acs.energyfuels.3c01473} {\path{doi:10.1021/acs.energyfuels.3c01473}}.

\bibitem{JMReport}
S.~Brown, L.~Cole, A.~Cowley, M.~Fujita, N.~Girardot, S.~Grant, J.~Jiang, R.~Raithatha, M.~Ryan, B.~Tang, Pgm market report, \url{https://matthey.com/documents/161599/404086/PGM+Market+Report+May23.pdf/2f048a72-74a8-8b23-f18e-c875000ed76b?t=1684144507321} (2023).

\bibitem{MINKE202123581}
C.~Minke, M.~Suermann, B.~Bensmann, R.~Hanke-Rauschenbach, \href{https://www.sciencedirect.com/science/article/pii/S0360319921016219}{Is iridium demand a potential bottleneck in the realization of large-scale pem water electrolysis?}, International Journal of Hydrogen Energy 46~(46) (2021) 23581--23590.
\newblock \href {https://doi.org/https://doi.org/10.1016/j.ijhydene.2021.04.174} {\path{doi:https://doi.org/10.1016/j.ijhydene.2021.04.174}}.
\newline\urlprefix\url{https://www.sciencedirect.com/science/article/pii/S0360319921016219}

\bibitem{CLAPP2023114140}
M.~Clapp, C.~M. Zalitis, M.~Ryan, Perspectives on current and future iridium demand and iridium oxide catalysts for pem water electrolysis, Catalysis Today 420 (2023) 114140.
\newblock \href {https://doi.org/https://doi.org/10.1016/j.cattod.2023.114140} {\path{doi:https://doi.org/10.1016/j.cattod.2023.114140}}.

\bibitem{IEAData}
IEA, Hydrogen production and infrastructure projects database, \url{https://www.iea.org/data-and-statistics/data-product/hydrogen-production-and-infrastructure-projects-database}, accessed: 2024-01-16.

\bibitem{REKSTEN202238106}
A.~H. Reksten, M.~S. Thomassen, S.~Møller-Holst, K.~Sundseth, \href{https://www.sciencedirect.com/science/article/pii/S0360319922040253}{Projecting the future cost of pem and alkaline water electrolysers; a capex model including electrolyser plant size and technology development}, International Journal of Hydrogen Energy 47~(90) (2022) 38106--38113.
\newblock \href {https://doi.org/https://doi.org/10.1016/j.ijhydene.2022.08.306} {\path{doi:https://doi.org/10.1016/j.ijhydene.2022.08.306}}.
\newline\urlprefix\url{https://www.sciencedirect.com/science/article/pii/S0360319922040253}

\bibitem{Babic_2017}
U.~Babic, M.~Suermann, F.~N. Büchi, L.~Gubler, T.~J. Schmidt, \href{https://dx.doi.org/10.1149/2.1441704jes}{Critical review—identifying critical gaps for polymer electrolyte water electrolysis development}, Journal of The Electrochemical Society 164~(4) (2017) F387.
\newblock \href {https://doi.org/10.1149/2.1441704jes} {\path{doi:10.1149/2.1441704jes}}.
\newline\urlprefix\url{https://dx.doi.org/10.1149/2.1441704jes}

\bibitem{CARMO20193450}
M.~Carmo, G.~P. Keeley, D.~Holtz, T.~Grube, M.~Robinius, M.~Müller, D.~Stolten, \href{https://www.sciencedirect.com/science/article/pii/S0360319918339338}{Pem water electrolysis: Innovative approaches towards catalyst separation, recovery and recycling}, International Journal of Hydrogen Energy 44~(7) (2019) 3450--3455.
\newblock \href {https://doi.org/https://doi.org/10.1016/j.ijhydene.2018.12.030} {\path{doi:https://doi.org/10.1016/j.ijhydene.2018.12.030}}.
\newline\urlprefix\url{https://www.sciencedirect.com/science/article/pii/S0360319918339338}

\bibitem{Sparber}
W.~Sparber, W.~Weiss, B.~Sanner, L.~Angelino, M.~D. Gregorio, N.~Fevrier, W.~Haslinger, A.~Kujbus, S.~Landolina, G.~Stryi-Hipp, W.~Helden, Clean hydrogen joint undertaking: Strategic research and innovation agenda 2021–2027, Clean Hydrogen Partnership (2021).

\bibitem{YU2020118194}
H.~Yu, L.~Bonville, J.~Jankovic, R.~Maric, \href{https://www.sciencedirect.com/science/article/pii/S0926337319309415}{Microscopic insights on the degradation of a pem water electrolyzer with ultra-low catalyst loading}, Applied Catalysis B: Environmental 260 (2020) 118194.
\newblock \href {https://doi.org/https://doi.org/10.1016/j.apcatb.2019.118194} {\path{doi:https://doi.org/10.1016/j.apcatb.2019.118194}}.
\newline\urlprefix\url{https://www.sciencedirect.com/science/article/pii/S0926337319309415}

\bibitem{autocatalysts}
M.~N.~. Insights, The outlook for platinum automotive demand, \url{https://market-news-insights-jpx.com/ose/commodities/article005405/}, [Online; accessed 28.03.2024] (2023).

\bibitem{streiks}
S.~A. Minerals~Council, Comprehansive facts and figure, 2013/14, \url{https://www.mineralscouncil.org.za/industry-news/publications/facts-and-figures}, [Online; accessed 28.03.2024] (2014).

\bibitem{IEAbattery}
P.~IEA, Global ev outlook 2023, \url{https://www.iea.org/reports/global-ev-outlook-2023}, [Online; accessed 19.12.2024] (2023).

\bibitem{CopperFoil}
L.-L. Lu, Z.-D. Wang, Q.~lu, K.-X. Song, H.-T. Liu, Y.-J. Zhou, F.~Zhou, Y.-M. Zhang, B.~Yang, Q.-Q. Zhu, W.-W. Lu, Advances in electrolytic copper foils: fabrication, microstructure, and mechanical properties, Rare Metals (2024) 1--36\href {https://doi.org/10.1007/s12598-024-02965-6} {\path{doi:10.1007/s12598-024-02965-6}}.

\bibitem{Umicore}
Umicore, Iridium, \url{https://pmm.umicore.com/de/preise/iridium/}, [Online; accessed 28.03.2024] (2024).

\bibitem{Forecast}
Forecasting: Principles and practice, \url{https://otexts.com/fpp2/}, accessed: 2024-02-10.

\bibitem{Pt}
A.~M. Portfolios, Global platinum primary mine supply plus recycling and demand estimates, \url{https://auctusmetals.com}, [Online; accessed 18.07.2024] (202s).

\bibitem{REVERDIAU202139195}
G.~Reverdiau, A.~{Le Duigou}, T.~Alleau, T.~Aribart, C.~Dugast, T.~Priem, \href{https://www.sciencedirect.com/science/article/pii/S0360319921037022}{Will there be enough platinum for a large deployment of fuel cell electric vehicles?}, International Journal of Hydrogen Energy 46~(79) (2021) 39195--39207.
\newblock \href {https://doi.org/https://doi.org/10.1016/j.ijhydene.2021.09.149} {\path{doi:https://doi.org/10.1016/j.ijhydene.2021.09.149}}.
\newline\urlprefix\url{https://www.sciencedirect.com/science/article/pii/S0360319921037022}

\bibitem{WPIC2025Q4}
{World Platinum Investment Council}, \href{https://platinuminvestment.com/files/328571/WPIC_Platinum_Quarterly_Q4_2024.pdf}{Platinum quarterly q4 2024}, Report, prepared by Metals Focus; provides annual refined platinum mine production and supply-demand analysis (Mar. 2025).
\newline\urlprefix\url{https://platinuminvestment.com/files/328571/WPIC_Platinum_Quarterly_Q4_2024.pdf}

\bibitem{EIKENG2024433}
E.~Eikeng, A.~Makhsoos, B.~G. Pollet, \href{https://www.sciencedirect.com/science/article/pii/S036031992401783X}{Critical and strategic raw materials for electrolysers, fuel cells, metal hydrides and hydrogen separation technologies}, International Journal of Hydrogen Energy 71 (2024) 433--464.
\newblock \href {https://doi.org/https://doi.org/10.1016/j.ijhydene.2024.05.096} {\path{doi:https://doi.org/10.1016/j.ijhydene.2024.05.096}}.
\newline\urlprefix\url{https://www.sciencedirect.com/science/article/pii/S036031992401783X}

\bibitem{doi:10.1021/acs.jpclett.6b00216}
A.~Kongkanand, M.~F. Mathias, \href{https://doi.org/10.1021/acs.jpclett.6b00216}{The priority and challenge of high-power performance of low-platinum proton-exchange membrane fuel cells}, The Journal of Physical Chemistry Letters 7~(7) (2016) 1127--1137, pMID: 26961326.
\newblock \href {http://arxiv.org/abs/https://doi.org/10.1021/acs.jpclett.6b00216} {\path{arXiv:https://doi.org/10.1021/acs.jpclett.6b00216}}, \href {https://doi.org/10.1021/acs.jpclett.6b00216} {\path{doi:10.1021/acs.jpclett.6b00216}}.
\newline\urlprefix\url{https://doi.org/10.1021/acs.jpclett.6b00216}

\bibitem{GIBSON2023108216}
B.~A. Gibson, G.~Nwaila, M.~Manzi, Y.~Ghorbani, S.~Ndlovu, J.~Petersen, \href{https://www.sciencedirect.com/science/article/pii/S0892687523002303}{The valorisation of platinum group metals from flotation tailings: A review of challenges and opportunities}, Minerals Engineering 201 (2023) 108216.
\newblock \href {https://doi.org/https://doi.org/10.1016/j.mineng.2023.108216} {\path{doi:https://doi.org/10.1016/j.mineng.2023.108216}}.
\newline\urlprefix\url{https://www.sciencedirect.com/science/article/pii/S0892687523002303}

\end{thebibliography}





\end{document}